\documentclass[
 reprint,
 twocolumn,
 amsmath,
 amssymb,
 aps,
 prl,
 nofootinbib,
 tightenlines,
 nobibnotes
]{revtex4-2}

\usepackage{graphicx}
\usepackage{dcolumn}
\usepackage{bm}
\usepackage{hyperref}

\usepackage{mathrsfs,amsmath} 
\usepackage{stfloats}  

\DeclareMathOperator{\csch}{csch}

\usepackage{color}

\definecolor{Green}{RGB}{0,100,0}
\definecolor{Blue}{RGB}{51,153,255}
\definecolor{Red}{RGB}{151,010,010}

\begin{document}

\title{Amplitude Structure of Optical Vortices Determines Annihilation Dynamics}

\author{Jasmine M. Andersen, Andrew A. Voitiv, Patrick C. Ford,}
\author{Mark E. Siemens}%
\email[]{msiemens@du.edu}%
\affiliation{$^1$Department of Physics and Astronomy, University of Denver, Denver, CO 80208, USA
}%

\begin{abstract}
We show that annihilation dynamics between oppositely charged optical vortex pairs can be manipulated by the initial size of the vortex cores, consistent with hydrodynamics. When sufficiently close together, vortices with strongly overlapped cores annihilate more quickly than vortices with smaller cores that must wait for diffraction to cause meaningful core overlap. Numerical simulations and experimental measurements for vortices with hyperbolic tangent cores of various initial sizes show that hydrodynamics governs their motion, and reveal distinct phases of vortex recombination; decreasing the core size of an annihilating pair can prevent the annihilation event. 
\end{abstract}

\maketitle

Vortices and their dynamics have fascinated scientists for centuries because of their prevalence in both classical and quantum mechanical systems~\cite{milne1996theoretical, gessner2019imaging, groszek2018motion, nilsen2006velocity, haldane1985quantum}. Significant efforts have been made particularly in studying optical vortices, with a subset of the work dedicated to quantifying optical vortex dynamics in laser beams, including work focused on vortex pairs~\cite{indebetouw1993optical, rozas1997propagation, roux1995dynamical, roux2004coupling}. Recent work also showed that by treating light as a two-dimensional compressible fluid, the dynamics of optical vortices can be fully understood by including vortex ellipticity in the hydrodynamic models~\cite{andersen2021tilt}. However, even without considering vortex ellipticity, changes in the background field gradients impact vortex dynamics. For example, by adding an additional phase gradient to a Gaussian field containing an oppositely charged vortex pair, the annihilation rate is accelerated~\cite{chen2008accelerating}. The dynamics of same-charge optical vortex pairs with small or large core structure were also compared, revealing different dynamical results~\cite{rozas1997experimental,rozas1997propagation}. The impact of changing the
initial size of the vortices in this way for an oppositely charged
pair and the direct impact of these changes to the initial field on annihilation events or dynamics has yet to be considered. 

In this letter, we use this initial amplitude structure of the vortex cores to modify oppositely charged vortex pair dynamics within a linear optical fluid. In particular, we employ the concept of hydrodynamics for laser light \cite{andersen2021tilt}, in which vortex velocity is governed by the gradients of phase and amplitude of the background fluid. This work demonstrates that the amplitude gradient is an important mechanism for facilitating annihilation events, which leads us to specifically use this degree of freedom to control the annihilation rate of these vortex pairs. We first numerically propagate a variety of initial condition fields to predict the vortex motion as a function of propagation. We find that the linearly shaped (large) cores annihilate more quickly than the hyperbolic tangent cores with a small radius and attribute this to the position at which a vortex is able to access the amplitude gradient from a neighboring vortex via diffraction of the core. The change in annihilation distance with initial core size is confirmed experimentally. We also numerically consider the annihilation condition based on initial separation and size of the cores. We find that the rate of annihilation can be controllably sped up or slowed down (or even prevented) by appropriate modification to these two parameters.

Consider an initial field given by the expression
\begin{equation}
     \psi(r,\phi,0) = \psi_{\mathrm{vortex}_-}\psi_{\mathrm{vortex}_+}\psi_{\mathrm{host}} = A_{-} e^{-i \phi_-}  A_+ e^{i \phi_+} e^{-r^2/w_0^2},
     \label{eq:psitwovortex}
\end{equation}
\noindent where the subscripts are related to the $\pm1$ vortex charge, $A_\pm$ denotes the radial vortex amplitude profile, $\phi_\pm = \arctan\left( y/(x\mp x_0)\right)$ describes the phase of the each vortex, $r=\sqrt{x^2+y^2}$, and $w_0$ is the host beam waist. The two opposite-charge vortices are placed symmetrically along the $x$-axis at locations $x=\pm x_0$ such that the initial separation between the vortices is $\mathrm{v}_{\mathrm{s}} = 2x_0$, with the positively (negatively) charged vortex on the right (left). Each vortex amplitude profile is taken to have a hyperbolic tangent shaped core such that
\begin{equation} \label{tanh}
A_{\pm}(r,\phi) = \tanh\left( \frac{\sqrt{\left(x \mp x_0\right)^2 + y^2}}{\mathrm{c_r}} \right).
\end{equation}
The motivation for choosing a hyperbolic tangent function is that for very small values of core radius, $\mathrm{c_r}$, the core approaches a delta function or point-core amplitude and for large values of $\mathrm{c_r}$, the core approaches the large, linear core limit. This allows us to scale the overlap (or lack thereof) of the amplitudes of each vortex at the $z=0$ initial condition. For unit charge vortices, only linear core amplitude structures are ``stable'' in free space: i.e., Laguerre-Gaussian modes for $\pm 1$ vortices. Thus, upon propagation from $z = 0$, ``tanh'' core vortices will diffract until they have linear cores. Presumably, the process of diffraction of these tanh cores will result in an altered trajectory for a given vortex pair.

\begin{figure}[h!]
\centering
\includegraphics[width=\linewidth]{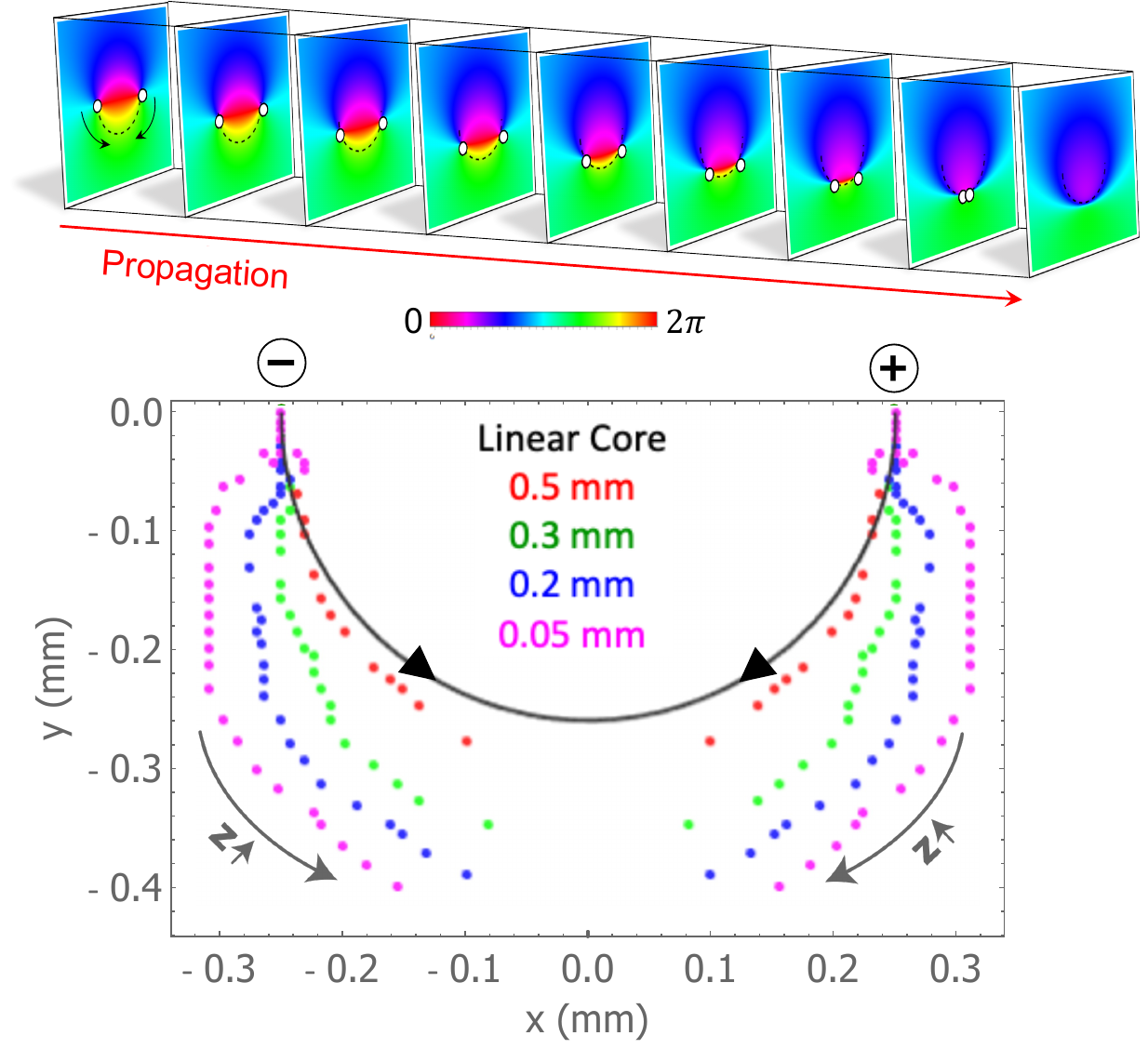}
\caption{(Top) Slices of the amplitude and phase for the evolving field of oppositely-charged vortices with linear ($\mathrm{c_r}\gg x_0$) cores are shown. (Bottom) Trajectories in the $xy$-plane for values of $c_r$ ranging from linear to very small are plotted. Each point represents a successive, equal step in $z$.}
\label{fig:numerical}
\end{figure}

Summarized in Fig. \ref{fig:numerical}, we use numerical simulation to anticipate such trajectories for a variety of oppositely-charged vortex pairs with distinct initial core radii, but equivalent beam waist $w_0=20$ mm ($w_0\gg x_0$) and separation $x_0=0.25$ mm. The numerical field is calculated at incremental $z$-steps via the angular spectrum method where the propagated field as a function of $z$ can be found via
\begin{equation}
    \psi(x,y,z)= \mathscr{F}^{-1}\left\{\mathscr{F}\{\psi(x,y,0)\}*H(f_x,f_y)\right\},
\end{equation}
where $H(f_x,f_y)=e^{\imath k z}e^{-\imath \pi \lambda z (f_x^2+f_y^2)}$, for wavelength $\lambda$ and spatial frequencies ($f_x$, $f_y$), is the transfer function in the paraxial approximation (hence, we employ a Fresnel propagation algorithm)~\cite{goodman2005introduction}. With the propagated field, the real and imaginary zero intersections locate the vortices at each step which are then used to plot the vortex trajectories.

The top row of Fig.~\ref{fig:numerical} shows the simulated phase contours at increasing propagation distances up until the annihilation point for a vortex pair, where the core size is very large compared to the vortex separation (i.e. $\mathrm{c_r}\gg x_0$). The result is a half-circle trajectory in the transverse plane, plotted as a solid black line in the bottom of Fig. \ref{fig:numerical}, consistent with prior work on linear core vortex pairs~\cite{indebetouw1993optical}. With this verification, we then compiled the trajectories for several pairs with decreasing core sizes. These trajectories, tracked up to the annihilation point, are also plotted in the bottom of Fig.~\ref{fig:numerical} for comparison.

As the cores decrease in size, two observations from the plot can be made: (1)  the trajectories increasingly deviate from the linear core half circle in the $xy$-plane and (2) the beam must propagate a farther distance before the vortices reach the point of annihilation, indicated by the larger number of $z$-steps. We can interpret these results using what is known about the impact of background phase and amplitude gradients on vortex motion. The phase gradient from the right, positive vortex at the location of the left vortex is initially downward via the right hand rule causing the left vortex to initially move downward. Similarly, the initial phase gradient from the left, negative vortex at the location of the right vortex is also in the downward direction, causing the right vortex to also move downward. Additionally, the contribution to the velocity on a given vortex from the background Gaussian amplitude, and the core of the other vortex is initially in the ($-\hat{y}$) direction \cite{andersen2021tilt}. This explains the initial vertical ($-\hat{y}$) motion of the vortices. In the presence of no amplitude gradients at all, the vortices would simply follow straight-line trajectories, as expected for an incompressible fluid system~\cite{milne1996theoretical}; for a single vortex in a Gaussian, the vortex also follows a straight line trajectory because the background gradients balance in such a way that the ($\hat{x}$) components cancel \cite{rozas1997propagation,andersen2021tilt}. For small cores with no initial overlap, we see this straight downward motion for small propagation distances which is consistent with prior work interpreting vortex motion as ``fluid-like" in the region very near to $z=0$~\cite{rozas1997experimental}. Another way to understand this is to say when the cores are very small at z=0, they initially move as if they are alone in the beam.

\begin{figure*}[h!]
\centering
\includegraphics[width=\linewidth]{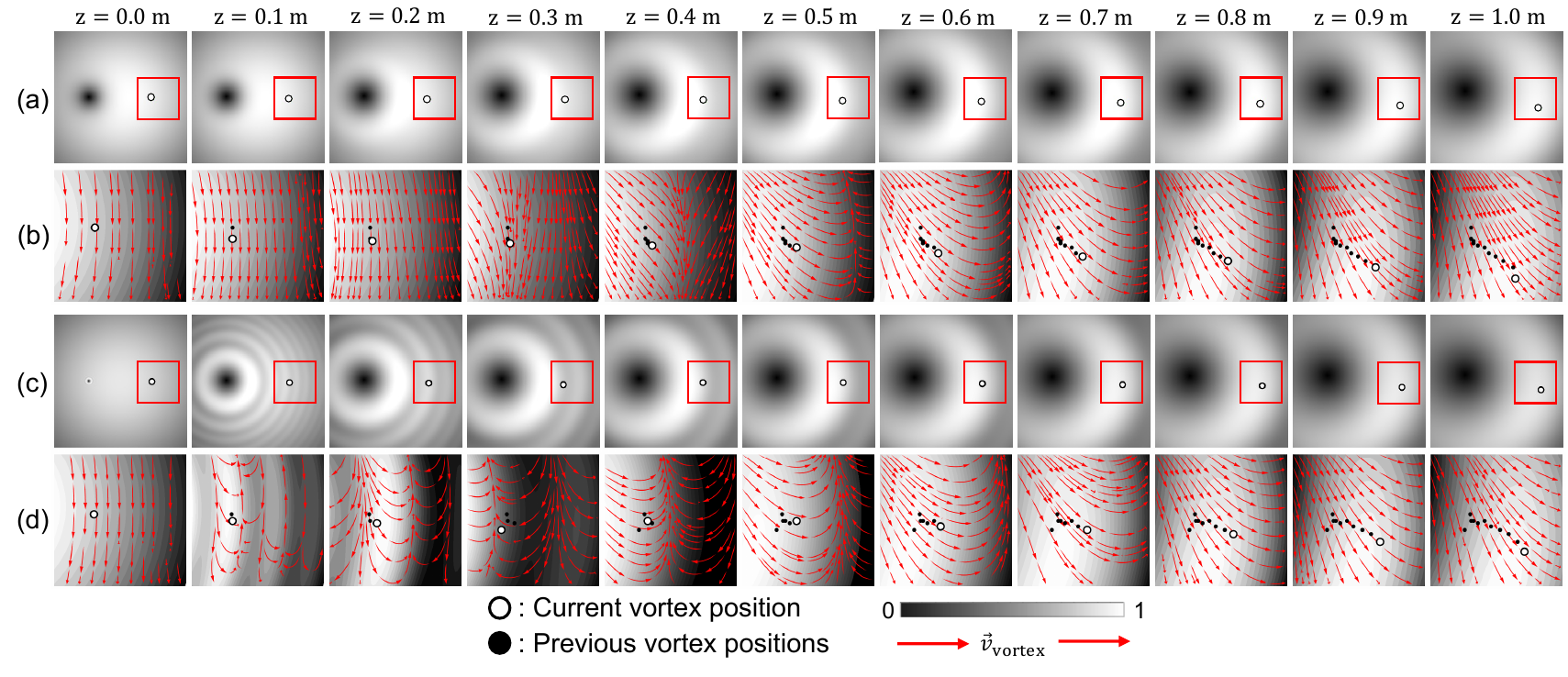}
\caption{(a) Simulated plots of amplitude with propagation of a single vortex in a Gaussian with a wavelength of $633$ nm and initial parameters $w_0=1$ mm, $x_0=0.30w_0$, and $\mathrm{c_r}=0.15w_0$ are shown with the marked location of the right vortex (black circle) from the measured two-vortex trajectory. (b) Plots of the highlighted regions from (a) along with red velocity vectors calculated via the untilted vortex velocity equation, given in Eqn. \ref{velocity} \cite{andersen2021tilt}, are shown. Black dots mark the prior vortex locations at each step for comparison of the actual trajectory with the velocity prediction. Vortex positions are measured by calculating the intersections of real and imaginary zeroes in the field. (c-d) show the same as in (a) and (b) with a wavelength of $633$ nm and initial parameters $w_0=1$ mm, $x_0=0.30w_0$, and $\mathrm{c_r}=0.005w_0$.}
\label{fig:diffractingcores}
\end{figure*}

However, small cores quickly diffract, resulting in contributions to the vortex velocity from the background core amplitude gradients at later propagation distances where one vortex is no longer effectively alone in the beam. To conceptually understand the impact of these gradients on the vortex motion, we must look at the changes in the background field with $z$. For our purposes, as shown in Fig. \ref{fig:diffractingcores}, we use an approximation of the background field which consists of only one vortex in the Gaussian beam and compare that with the motion of the other vortex. This will not be the true background field, since to find the background field at any given propagation distance one would need to divide out the vortex of interest, including its ellipticity, from the field. Additionally, we know that the evolving ellipticity of the vortex itself can influence the trajectories, as discussed in~\cite{andersen2021tilt}. However, using this approximation at early propagation times where the vortices are expected to be mostly circular demonstrates the connection between the differing dynamics based on a different sized core. Particularly, the outward motion of a given vortex seen in Fig.~\ref{fig:numerical} can be attributed to the arrival of the diffraction wave from the neighboring vortex.

In Fig.  \ref{fig:diffractingcores}, we show the numerically propagated initial field, given the assumptions mentioned in the previous paragraph, and calculate the anticipated untilted vortex velocity of the right vortex. The background field is set with the same parameters as in the two vortex field, and field gradients are calculated using a two-point finite difference method~\cite{sauernumerical}.  In the case of row (a), the diffraction mainly comes in one wave that pushes the vortex outward upon its arrival. In row (c) with the very small core, ringing in the diffraction field, also seen in ~\cite{rozas1997propagation}, becomes stronger and successive waves create a temporary spiral-like motion of the vortex, followed by the outward motion from the last diffracting peak. These investigations could possibly be extended to quantify the change in momentum experienced by the vortex surfing on these radial rings (including a delineation of the separate effects from the background phase and amplitude contributions). In both cases, rows (a) and (c), despite making the approximations mentioned, the calculated vortex velocity   is a reasonable prediction of the vortex motion and shows the impact of the diffracting core on the neighboring vortex. In the case of the linear core, the absence of the outward motion can be seen as a result of there being no diffraction wave from the other vortex.

\begin{figure}[h!]
\centering
\includegraphics[width=\linewidth]{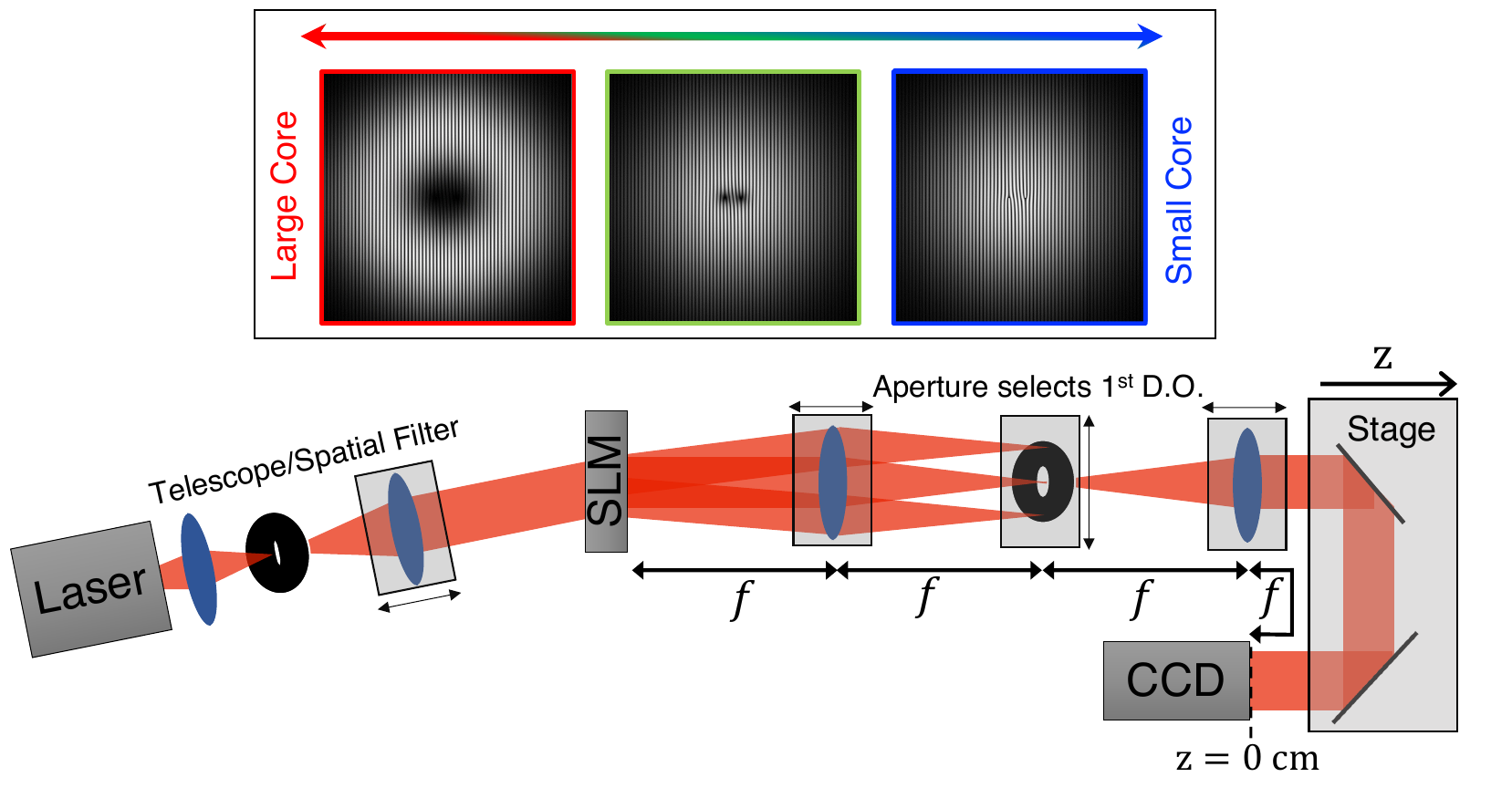}
\caption{A schematic of the experiment shows a HeNe laser incident on a spatial light modulator that generates an oppositely charged vortex pair. Holograms for three different core sizes are shown in the top row. A $4f$ imaging system collects the light and a translation stage adjusts the path length before the light is recorded on a CCD. }
\label{fig:setup}
\end{figure}

In addition to our simulations, we experimentally measured the trajectories for a few cases. For the experiment, shown in Fig.  \ref{fig:setup}, a HeNe laser is incident on a spatial light modulator that projects holograms that contain diffraction gratings made by the interference of a plane wave and the initial field given by Eqn. \ref{eq:psitwovortex}, such that this field is produced in the first diffracted order of the transmitted light. The SLM works in amplitude mode, and the gratings are modulated in both amplitude and phase to create the output beam. The first diffracted order is then spatially filtered at the focus of a 4-$f$ imaging system by an aperture. All other diffracted orders are therefore blocked. The beam is then directed to a translation stage with a retroreflector that sends the beam to a Wincam LCM CCD. The CCD is placed at the imaging plane and the evolution of the field with propagation is measured by moving the stage to increase the path length before the CCD. 

At each step in $z$, we used collinear phase shifting digital holography to measure the full complex field~\cite{andersen2019characterizing}. This process entails compiling five images recorded from each $z$-step: one intensity image and four images of the signal interfered with the reference beam at different phase steps used to calculate the phase. A signal power to reference power ratio of 0.95:0.05 is used in the holograms to obtain the cleanest field measurement at the very dark centers of the vortices. Additionally, we measured a single centered vortex (an $\ell = 1$ Laguerre Gaussian mode) to center and crop the data to remove the effects from any external drifts in the system. A set of single, centered vortices with $\ell=\pm 10$ like this were also used to carefully align the co-axial position and angle of the SLM such that the diffracted order is perpendicular to the SLM before commencing on experiments with dipoles; this ensured the intended symmetrical displacement of dipoles about the center of the Gaussian. Once the experimental field is measured and calibrated, we used the same method of intersecting real and imaginary zeros to locate and track the vortices. Everything is then in place to quantitatively analyze the trajectory data. 

\begin{figure}[h!]
\centering
\includegraphics[width=\linewidth]{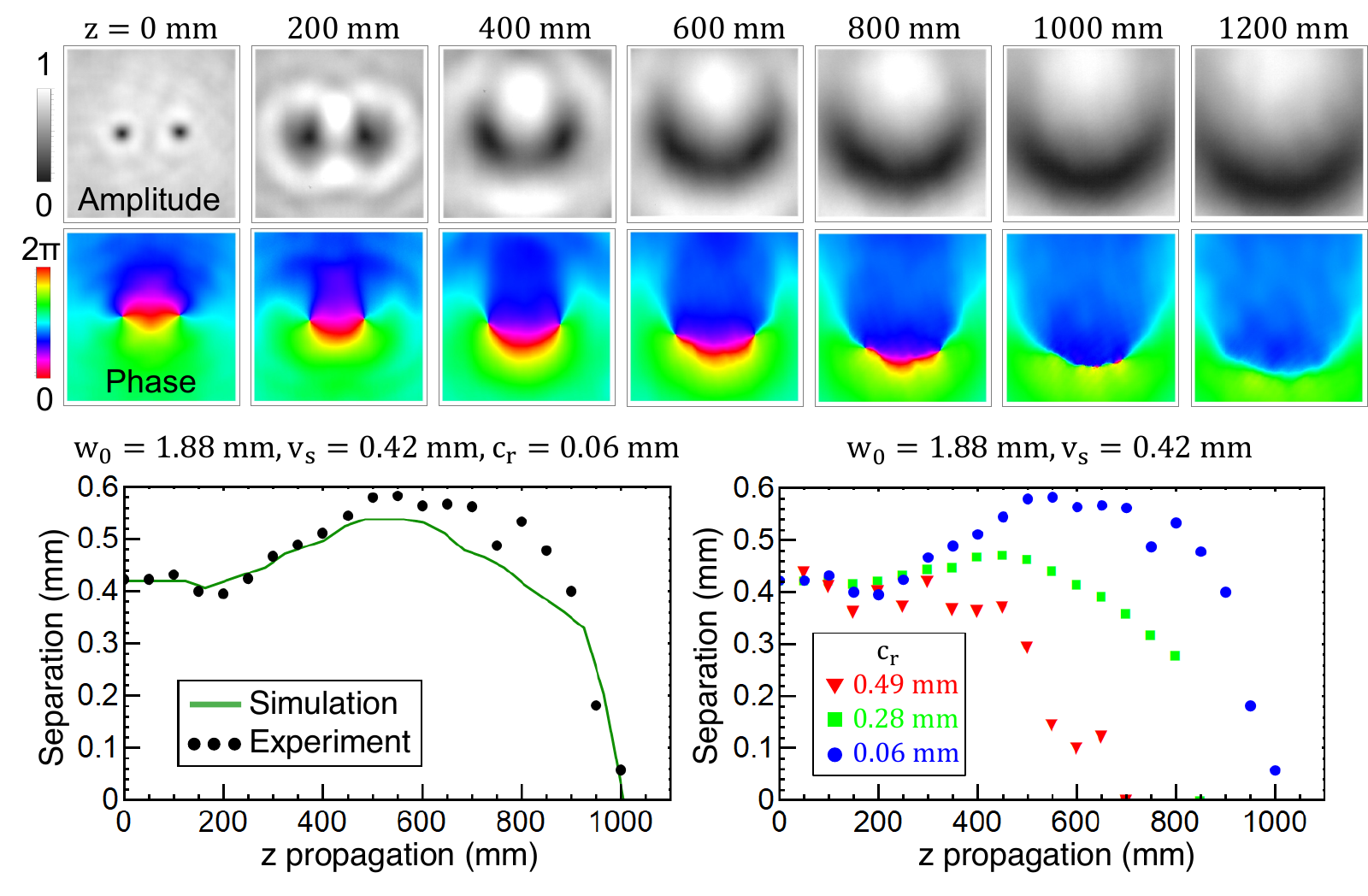}
\caption{(Top) Experimental slices of the field are shown for a set of propagation steps. (Bottom Left) Comparison of vortex separation as a function of propagation shows good agreement between simulation and experiment. (Bottom Right) A set of measurements for three different sized vortex cores using holograms, such as those found in Fig.  \ref{fig:setup}, verifies that larger cores annihilate more quickly than small cores.
}
\label{fig:separation}
\end{figure}

We used a 2D fitting routine, with Equation~\ref{eq:psitwovortex}, on the amplitude of the field at the imaging plane to measure the initial condition parameters including the beam waist, initial vortex separation and core sizes which were then used in the simulation for comparison with the experimental measurement. In the bottom left of Fig.  \ref{fig:separation}, we show a quantitative comparison of the separation between the vortices. The experimental vortex separation and the numerical simulation also show a very strong agreement between the experiment and simulation, including the fact that the vortices are pushed farther away from each other before moving toward each other and annihilating. The vortices are not small enough to observe the small oscillations shown in the $\mathrm{c_r} =0.05$ mm case in Fig.  \ref{fig:numerical} since the setup is limited by the numerical aperture of the system. It is possible to use lenses of smaller focal lengths, but the trade off was made by using longer focal length lenses to ease in alignment of the imaging system, reducing the likelihood of magnification errors, and therefore inaccurate vortex separation measurements, with beam propagation. Additionally, we experimentally measured vortex separation for three different core sizes, shown in the bottom right of Fig.~\ref{fig:separation}, which confirm smaller annihilation distances for large cores and farther distances for small cores. The top row of Fig. \ref{fig:separation} is experimental data with no fitting or approximations applied; likewise, in the bottom row, all experimental vortex locations (dots and symbols) are measured directly from the experimental data as measured by the CCD---not from fitted fields. These identified positions of the vortices are overlaid with the corresponding experimental phase measurement, to verify that the vortex positions identified are consistent to within one to two pixels of the vortex center of the phase image.

Lastly, we numerically consider a variety of initial conditions to interrogate annihilation events based on initial beam parameters. As can be seen in the top row of Fig. \ref{fig:separation}, for vortices that annihilate, the dynamics resemble those found in Fig. \ref{fig:numerical}. Initial conditions can be applied such that the rate of annihilation is slowed down, as shown in the bottom of Fig. \ref{fig:separation}, or even prevented and/or driven to occur unexpectedly. On this last point, even for ``point cores'' ($\mathrm{c_r} \rightarrow 0$), we identified a critical vortex separation of $x_{0,\mathrm{crit}} \approx 0.354 w_0 = w_0 / \sqrt{8}$ as the onset of annihilation.

\begin{figure}[h!]
\centering
\includegraphics[width=\linewidth]{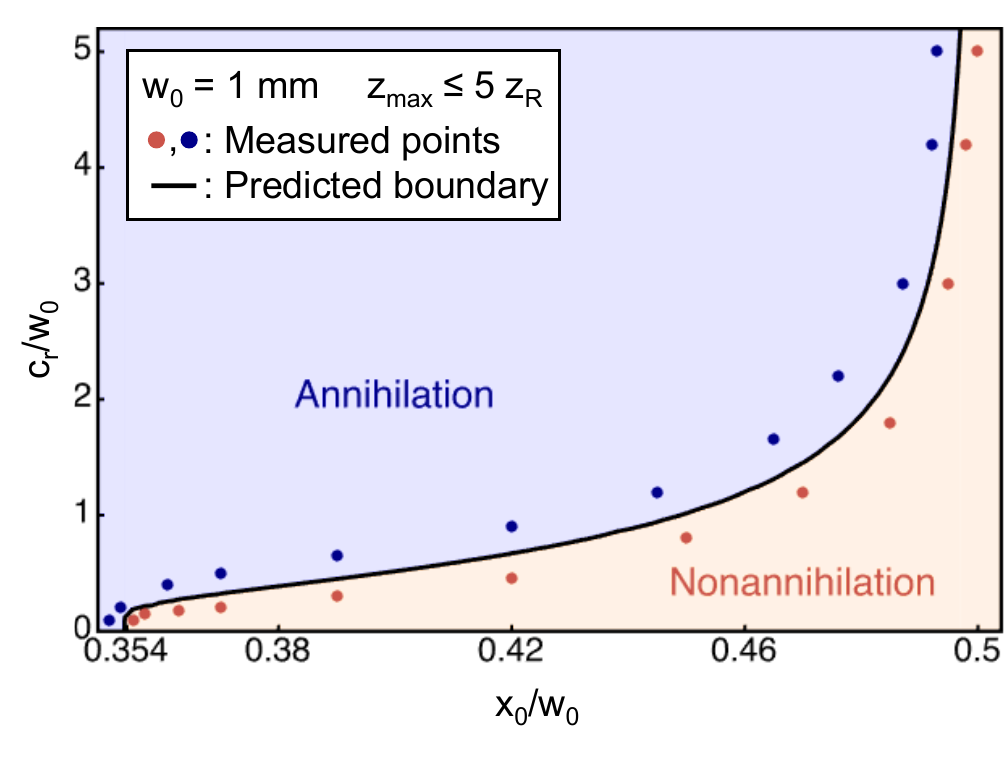}
\caption{Phase diagram of vortex dipole annihilation/nonannihilation, based on the initial parameters of core size ($\mathrm{c_r}$) and vortex separation ($x_0$). Measured points are numerical simulations, in which a dipole is propagated (no further than $5$ Rayleigh lengths, $z_R$) and observed to either annihilate or not---determined by the two parameters on the axes. The measurements closely obey the predicted phase boundary of Eqn. \ref{boundary}, the black curve.}
\label{fig:phasediagram}
\end{figure}

Fig. \ref{fig:phasediagram} is a phase diagram, showing two distinct regions for which vortices either annihilate or do not, depending on the two parameters of \textit{initial}  vortex separation and core size. Numerically measured points---represented by blue and orange dots on Fig. \ref{fig:phasediagram}---represent pairs of initial parameters ($x_0,\mathrm{c_r}$) which either led to vortex annihilation or not within a distance of five Rayleigh lengths of beam propagation. A scale of uncertainty in the sensitivity of our numerical investigation is achieved by identifying two nearby points of both annihilation and nonannihilation, each chosen to straddle the boundary between the two regions.

The black curve in Fig. \ref{fig:phasediagram} is a predicted boundary based upon the hydrodynamic interpretation of laser light for the velocity of a single vortex \cite{andersen2021tilt}:
\begin{equation} \label{velocity}
    \vec{v}_{\mathrm{vortex}}=k \, \nabla \phi_{\mathrm{bg}}-\vec{k} \times \nabla \ln \rho_{\mathrm{bg}},
\end{equation}
where $\phi_{\mathrm{bg}}$ is the background phase from the other vortex, using $\phi_{\pm}$ defined earlier, and $\rho_{\mathrm{bg}}$ is the background amplitude from both the host Gaussian beam and the other vortex, $A_{\pm} \, e^{-r^2 / w_0^2}$, for $A_{\pm}$ defined in Eqn. \ref{tanh}. Using $\phi_{\mathrm{bg}}$ and $\rho_{\mathrm{bg}}$ as defined above, Eqn. \ref{velocity} predicts an initial velocity of one of the vortices as:
\begin{multline} \label{applied}
    v_y(x_0, \mathrm{c_r})|_{x=x_0,y=0,z=0} = \\
    -\frac{2 \pi}{\lambda}\left\{ \frac{1}{2x_0} + \frac{2}{\mathrm{c_r}} \csch{\left( \frac{4x_0}{\mathrm{c_r}} \right)} - \frac{2x_0}{w_0^2} \right\},
\end{multline}
written for the right-side vortex of the dipole pair (the left-side vortex has an equivalent initial velocity, and both are initially in the $-\hat{y}$ direction only).

From Eqn. \ref{applied}, we seek a ``critical boundary''---pairs of initial parameters $x_0$ and $\mathrm{c_r}$ that would bisect the measurement outcomes of either annihilation or nonannihilation. To start, consider the linear core limit, $\rho_{\mathrm{bg}} \rightarrow (r + r_0) = (x + x_0)$, considering the right-hand vortex of a dipole on the $x$-axis. Application of Eqn. \ref{velocity} yields,

\begin{equation*}
    v_y|_{x=x_0,y=0,z=0} = -\frac{2 \pi}{\lambda} \left( \frac{1}{x_0} - \frac{2x_0}{w_0^2} \right).
\end{equation*}

For linear cores such as this, we can find the value of $x_0$ that separates outcomes of annihilation or nonannihilation. This amounts to setting the velocity above equal to a ``drift velocity'', where the vortices simply react to the phase gradient contribution from each other: $v_{\mathrm{drift}} = k \nabla \phi_{\mathrm{bg}}$.
\begin{equation*}
\begin{split}
    -\frac{2 \pi}{\lambda} \left( \frac{1}{x_0} - \frac{2x_0}{w_0^2} \right) &= -\frac{2\pi}{\lambda} \frac{1}{2x_0} \\ 
    \frac{1}{2x_0} - \frac{2x_0}{w_0^2} &= 0 \\
    \frac{x_0}{w_0} &= \frac{1}{2},
\end{split}
\end{equation*}
and indeed, Fig. \ref{fig:phasediagram} indicates that in the limit of $\mathrm{c_r} \gg w_0$, a separation of $x_0 = 0.5 w_0$ is the boundary between annihilation and nonannihilation.

For tanh-core vortices, we apply this same logic to Eqn. \ref{applied}. However, the ``drift velocity'' needs to be modified to account for the effects of the tanh-cores diffracting---as closely studied in Fig. \ref{fig:diffractingcores}. That is, there is inherently no ``ideal'' case where the tanh-core vortices will only react to phase gradients, because they will always surf on the diffraction waves with propagation. As a starting point, we considered increasing  the drift velocity by the tanh-core contribution, $2 \, k \, \mathrm{c_r}^{-1} \, \csch{\left( 4x_0 / \mathrm{c_r} \right)}$. But this term over-corrects the drift velocity; we need also to subtract its counter-part, the linear-core contribution $k (2x_0)^{-1}$. We define the new drift velocity for tanh-core vortices as the combination of the linear-core drift velocity and the arithmetic mean between these two corrections, leading to the equation
\begin{multline*}
    -\frac{2 \pi}{\lambda} \left[ \frac{1}{2x_0} -\frac{2x_0}{w_0^2} + \frac{2}{\mathrm{c_r}} \csch{\left( \frac{4x_0}{\mathrm{c_r}} \right)} \right] = \\
    -\frac{2 \pi}{\lambda} \left\{ \frac{1}{2x_0} + \frac{1}{2} \left[ \frac{2}{\mathrm{c_r}} \csch{\left( \frac{4x_0}{\mathrm{c_r}} \right) - \frac{1}{2x_0}} \right] \right\},
\end{multline*}
which is then simplified to
\begin{equation} \label{boundary}
    -\frac{2x_0}{w_0^2} + \frac{1}{\mathrm{c_r}} \csch{\left( \frac{4x_0}{\mathrm{c_r}} \right)} + \frac{1}{4x_0} = 0.
\end{equation}
In the limit of $\mathrm{c_r} \rightarrow 0$, this relation yields the critical separation distance $x_{0,\mathrm{crit}} = w_0 / \sqrt{8}$, as discussed above. The boundary in Fig. \ref{fig:phasediagram} is the zero-crossing line for the two-dimensional ($x_0,\mathrm{c_r}$) relation on the left-hand-side of Eqn. \ref{boundary}.

We see that the numerically recorded data points of Fig. \ref{fig:phasediagram} closely follow this predicted boundary from Eqn. \ref{boundary}.
It is surprising that just the initial condition \textit{alone} is able to predict the annihilation outcomes of a vortex dipole---and even more so that the numerical study follows the prediction so closely (within the sensitivity and propagation range chosen). But combined with the ``surfing'' dynamics of a vortex on the diffraction waves of its dipole-partner in Fig. \ref{fig:diffractingcores}, these findings show that effects of manipulating vortex structure can be intuited from the hydrodynamic interpretation of optical vortices in motion.

The results of Fig. \ref{fig:phasediagram} can be intuitively understood. For vortices with large cores that are spaced close together, annihilation occurs relatively quickly, matching the prediction that highly overlapped cores interact more strongly. For vortices that are spaced far enough apart and that have small enough cores, one would expect that the vortices would not have a sufficient overlap before the background Gaussian gradients dominate the vortex motion and keeps them apart for the entirety of the propagation. Consider a specific separation, for example  $x_0/w_0=0.46$ on the phase diagram, for which a linear core ($\mathrm{c_r} \gg w_0$) dipole is expected to annihilate. By simply reducing the size of the core to be $c_r/w_0 < 1$ for that separation, the annihilation event can be prevented. Conversely, though not immediately obvious, it makes intuitive sense that vortex pairs with sufficiently small separation will always annihilate, even with small cores, since the cores will need a much smaller propagation distance before fully overlapping. An example of this is $x_{0,\mathrm{crit}} = 0.354w_0$, as mentioned earlier.

To summarize, we have shown that vortex pair annihilation in linear optical systems can be tuned by simply changing the initial overlap of the vortex core functions. These changes can alter annihilation events, by speeding them up or slowing them down---or stopping annihilation altogether. This paper demonstrates that vortex annihilation events can be controlled by tuning the appropriate beam parameters, in accordance with laser hydrodynamics. This work could be extended to understand and control vortex interactions in more complex, many vortex systems with non-uniform core shapes and sizes using the same principles. Additionally, the insights here could potentially be applied to investigations of vortex annihilation events in other systems such as quantum fluids \cite{dominici2018, kwon2021}, particularly considering the number of previous studies of vortex dipoles in those settings \cite{neely2010, torres2011, kasamatsu2016, pshenichnyuk2018}.

\section*{Funding}

W. M. Keck Foundation and National Science Foundation (NSF, 1553905).

\section*{Disclosures}

The authors declare no conflicts of interest.

\section*{Data availability} 

Data underlying all results presented are available from the authors upon reasonable request.

\bibliographystyle{apsrev4-2}
\bibliography{references}

%

\end{document}